# The Pointing Error Due To Laser Propagation in Space Plasma for TianQin Gravitational Wave Detection

**Shenwei Zhou[1], Wei Su[1], Shiyuan Zhou[2], Chaofan Li[1], and Jixiang Zhang[1]**

[1]MOE Key Laboratory of TianQin Mission, School of Physics and Astronomy, TianQin Research Center for Gravitational Physics, Frontiers Science Center for TianQin, Gravitational Wave Research Center of CNSA, Sun Yat-Sen University (Zhuhai Campus), Zhuhai, China, [2]China Institute of Atomic Energy, Beijing, China

**Key Points:**

- The propagation of the laser in space plasma can induce pointing errors for gravitational wave detection projects, such as TianQin
- A ray-tracing based on the Runge-Kutta method is developed to assess the light deflection of the laser propagation in space plasma
- Light deflection in space plasma is negligible for TianQin, which indicates a positive outlook for the TianQin mission

**Correspondence to:**

W. Su,
suwei25@mail.sysu.edu.cn

**Author Contributions:**

**Conceptualization:** Wei Su
**Formal analysis:** Shenwei Zhou, Shiyuan Zhou
**Funding acquisition:** Wei Su
**Investigation:** Shenwei Zhou, Jixiang Zhang
**Methodology:** Shenwei Zhou, Shiyuan Zhou
**Project administration:** Wei Su
**Resources:** Wei Su, Jixiang Zhang
**Software:** Shenwei Zhou, Wei Su, Shiyuan Zhou, Jixiang Zhang
**Supervision:** Wei Su
**Validation:** Shenwei Zhou, Chaofan Li
**Visualization:** Shenwei Zhou, Wei Su
**Writing – original draft:** Shenwei Zhou
**Writing – review & editing:** Shenwei Zhou, Wei Su, Chaofan Li, Jixiang Zhang

## Abstract

TianQin, a space-borne gravitational wave observatory, is designed to detect gravitational-wave signals via laser interferometry. As the inhomogeneous space plasma surrounds the inter-satellite laser links, the propagation of the laser in space plasma can introduce pointing error noise. In this work, we quantitatively assess the effect of plasma on the pointing accuracy of TQ's inter-satellite laser links. We employ a magnetohydrodynamic model to simulate the plasma distribution around the TQ orbit for different orbital orientations. Using the fourth-order Runge-Kutta method to conduct the ray-tracing calculations over a TQ orbital cycle under a solar wind dynamic pressure close to the 23-year average, we find that the plasma-induced laser deflection angle and pointing eccentricity are on the order of $10^{-20}$ rad and picometers, respectively. These values are significantly lower than TQ's requirements. Our results demonstrate that the impact of space plasma on TQ's pointing error is negligible in moderate space weather conditions, indicating a positive outlook for the TQ mission.

## Plain Language Summary

TianQin is one of the space-based gravitational wave missions that is proposed to detect the gravitational wave sources by measuring the tiny changes in the distance between satellites using laser beams. This requires an extremely stable and precise measurement system. However, space is not a perfect vacuum, but is filled with a variety of media. Around the orbit of TianQin, the space plasma plays a role in affecting the measurement of inter-satellite lasers. A concern occurs whether space plasma could cause laser deflection enough to distort TianQin's measurements. Our study sets out to quantify this potential effect. By simulating the plasma environment around TianQin's orbit and meticulously calculating the path of the laser beams over time, we found that the plasma-induced deflection is incredibly small. The effect is orders of magnitude weaker than the mission's strict requirements. In that case, we conclude that the deflection of laser beams by space plasma is negligible. Our finding removes a potential concern, indicating a positive outlook for the TianQin mission.

## 1. Introduction

Gravitational waves played a decisive role in verifying Einstein's theory of relativity. In 2015, the Laser Interferometer Gravitational-Wave Observatory (LIGO) detected the first gravitational wave (GW) signal, marking the official beginning of a new era in GW astronomy (Abbott et al., 2016). Current GW detectors are mainly divided into three categories. One category includes ground-based detectors such as LIGO, along with the Advanced Virgo (Acernese et al., 2014) and KAGRA (Akutsu et al., 2019). Due to the influence of ground vibration noise and gravitational gradient noise, the lower limit of sensitivity for ground-based detectors is around 10 Hz (Aasi et al., 2015). However, GW sources in the frequency band between 0.1 and 10 Hz hold significant importance for astronomical and cosmological research. These sources include binary white dwarf systems (Amaro-Seoane et al., 2023; Huang et al., 2020), stellar-mass binary black holes (Caputo et al., 2020; Liu et al., 2020), as well as inspiral (Isoyama et al., 2018), merger, and ringdown events (Babak et al., 2017; Wang et al., 2019), among others. Therefore, to extend the detectable frequency band of GW, several space-based GW missions have been proposed. These include the Laser Interferometer Space Antenna (LISA), a collaborative project between the European Space Agency and the National Aeronautics and Space Administration (Amaro-Seoane et al., 2017; eLISA Consortium et al., 2013), the DECi-hertz Interferometer Gravitational Wave Observatory (DECIGO) in Japan (Kawamura et al., 2011), the Taiji (TJ) (Gong et al., 2015), and the TianQin (TQ) project (Luo et al., 2016; Li et al., 2025) programs led by China. Additionally, pulsar timing arrays (PTAs) play a role in the search for the GW background within the nanohertz frequency band (Agazie et al., 2023; Antoniadis et al., 2022; Chen et al., 2021; Xu et al., 2023). And the NANOGrav collaboration has released its 15-year data set, with evidence for the presence of Hellings-Downs correlations, providing key evidence for a GW (Agazie et al., 2023).

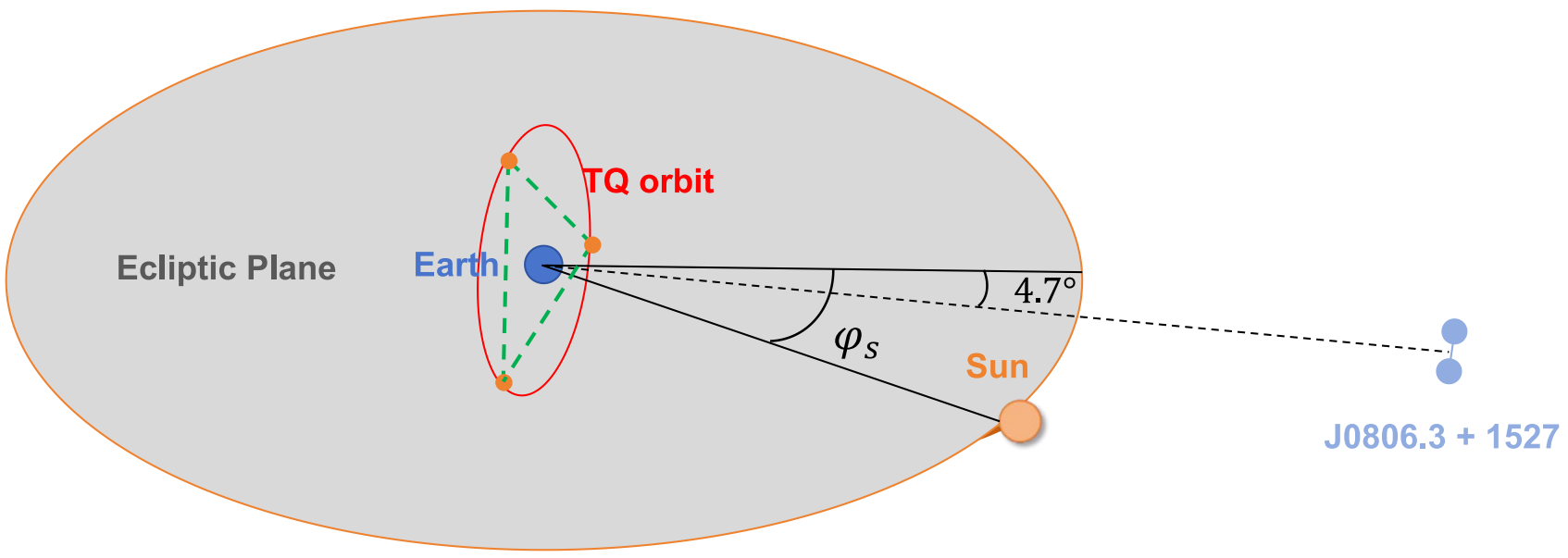

**Figure 1.** An illustration of the TQ satellite orbit. The Earth is set as the coordinate origin and remains stationary. The TQ orbit is marked by the red line, with the inter-satellite laser beams marked by green dashed lines. The black dashed line represents the normal vector of the TQ orbital plane, pointing toward the reference source J0806.3 + 1527. The TQ orbital plane forms a 4.7° angle with the ecliptic plane. The gray plane is the ecliptic plane on which the sun rotates in this coordinate system.

Among all three kinds of methods mentioned above, space-based GW detectors employ responsive inter-satellite laser interferometry, with satellite separations ranging from hundreds of thousands to millions of kilometers. For instance, TQ is designed to primarily detect GW in the frequency band of $10^{-4}$–1 Hz. The TQ constellation consists of three satellites forming an equilateral triangle orbiting the Earth at a radius of approximately $1 \times 10^5$ km. The separation between satellites is about $1.7 \times 10^5$ km, pointing toward the reference source J0806.3 + 1,527 (Luo et al., 2016; X.-C. Hu et al., 2018). LISA forms an equilateral triangle constellation with arm lengths of $2.5 \times 10^6$ km (Amaro-Seoane et al., 2017; Karnesis et al., 2024), and it operates in an Earth-trailing heliocentric orbit between 50 and 65 million km from Earth. With a nominal mission lifetime of 4 years, LISA aims to detect GW in the frequency range of $10^{-5}$–$10^{-1}$ Hz. The Taiji program, on the other hand, is designed to observe GW in the frequency band of $10^{-4} - 1$ Hz (Hu & Wu, 2017; Luo et al., 2020). It will operate in an Earth-leading heliocentric orbit, positioned approximately 20° ahead of Earth, and form an equilateral triangle constellation with an arm length of $3 \times 10^6$ km. The regions surrounding these GW detectors are not a perfect vacuum, but rather filled with numerous media such as high-energy particles (Han et al., 2023, 2024; Reames, 2013), space plasma, and magnetic fields (Su, 2025). Among these, the orbital formation configurations of LISA and Taiji around the Sun mean that they are primarily influenced by the solar wind environment. In contrast, TQ's geocentric orbit scheme will be subject to the combined effects of Earth's magnetosphere and the solar wind (Su, 2025).

LISA, Taiji, and TQ missions all employ lasers at a wavelength of 1064 nm for inter-satellite communication. Both LISA and Taiji operate with a laser power of 2 W and require a pointing stability of ≤10 nrad under direct current (DC) bias, with the point jitter to $\leq 10$ nrad/$\sqrt{\text{Hz}}$ (Amaro-Seoane et al., 2017; W.-R. Hu & Wu, 2017). TQ uses a higher laser power of 4 W and maintains the same pointing stability requirement of ≤10 nrad under DC bias, with the point jitter to $\leq 10$ nrad/$\sqrt{\text{Hz}}$ (Duan et al., 2021; Luo et al., 2016).

In this paper, Section 2 introduces the TQ Mission and its high precision requirements in details. Section 3 describes the methodology for obtaining the plasma distribution around the TQ orbits and explains how the inhomogeneous plasma along the inter-satellite links leads to laser deflections. Section 4 presents the variation of the laser deflection angle along the propagation path at a specific time instance, followed by the evolution of $\varepsilon$ and $\alpha$ over one full orbital cycle under varying satellite positions and space environment conditions. Section 5 carries out preliminary data processing and fundamental analysis of the data obtained in Section 4. Finally, we summarize the results and draw conclusions.

## 2. TianQin Mission and Its Measurement Requirements

TQ is a geocentric GW detector, the illustration is shown in Figure 1. The TQ satellite completes one orbit around the Earth every 3.65 days. The Earth is set as the coordinate origin. The gray plane represents the ecliptic plane.

The TQ orbit is marked by the red line, while the green dashed lines indicate the inter-satellite links. The black dashed line represents the normal vector of the TQ orbital plane, pointing toward the reference source J0806.3 + 1527, and it forms a 4.7° angle with the ecliptic plane. $\varphi_s$ is the acute angle between the projection of the normal direction of the TQ orbital plane on the ecliptic plane and the Sun-Earth line. Because of the small tilt angle of 4.7° between the normal of the TQ orbit plane and the Ecliptic plane, the TQ orbit plane is approximately perpendicular to the Ecliptic plane. Therefore, $\varphi_s$ reaches its minimum value of 0° and its maximum value of 90°. For $\varphi_s = 0°$ plane, the line of intersection between the orbital plane of TQ and the ecliptic plane is the local dawn-dusk line; For $\varphi_s = 90°$ plane, the line of intersection between the orbital plane of TQ and the ecliptic plane is the local noon-midnight line. Consequently, we choose three typical positions with $\varphi_s = 0°$, 45°, and 90° for our calculation. Moreover, according to Figure 1, the TQ orbital plane is considered motionless in the coordinate system. The Earth's orbital motion around the Sun is consequently represented by the Sun's annual rotation within this frame. This rotation causes the Sun-Earth line to revolve accordingly, resulting in a yearly periodic variation of $\varphi_s$. As the period of the TQ satellite movement around the Earth is about 3.65 days, the TQ satellite move around the Earth 100 cycles in 1 year, the precession of $\varphi_s$ in 1 cycle is small, and $\varphi_s$ is assumed to be approximately constant during 1 cycle in this work.

TQ is designed primarily to detect GW in the frequency band of $10^{-4} - 1$ Hz. Sources expected for TQ include the inspiral of Galactic ultra-compact binaries, and stellar-mass binary black holes, extreme mass ratio inspirals, among others (Luo et al., 2025; Mei et al., 2020). The sensitivity of TQ is also set to be capable of detecting GWs emitted by the reference source J0806.3 + 1527 (Luo et al., 2016), with a position measurement accuracy of $1\ \mathrm{pm}/\sqrt{\mathrm{Hz}}$ at ∼6 mHz and a residual acceleration accuracy of $10^{-15}\ \mathrm{m/s^2}/\sqrt{\mathrm{Hz}}$ at ∼6 mHz. This extremely high requirement leads to the assessment of the effects of multiple factors. Space magnetic fields can introduce acceleration noise (Su et al., 2020, 2023). The laser propagating in the space plasma can lead to optical path difference (OPD) noise for both geocentric and heliocentric GW detectors (Jennrich et al., 2021; Lu et al., 2021; Smetana, 2020; Su et al., 2021). Su et al. (2021) revealed that time-delay interferometry (TDI) can suppress OPD noise during the laser propagation in space plasma, and this finding was subsequently confirmed (Jing et al., 2022; Xie et al., 2024; Zhao et al., 2024). Besides the OPD noise caused by laser propagation in space plasma, the optical beam deflection effect resulting from laser propagation remains insufficiently investigated.

In the case of TQ, its inter-satellite laser links traverse regions such as the solar wind, bow shock, magnetopause, and magnetotail (Su et al., 2020, 2021). The plasma surrounding the links is an inhomogeneous medium influenced by both the solar wind and the geomagnetic field (Chen, 2011; Lu et al., 2019; Wang et al., 2020). It causes the laser beams to deflect while propagating through the plasma, generating plasma-induced deflections. Previous studies have shown that the OPD noise caused during the laser propagation in space plasma can reach the picometer ($10^{-12}$ m) level in time domain (Su et al., 2021), which is comparable to the measurement precision required by the TQ mission. Given this extreme sensitivity, it is necessary to examine whether other plasma-related effects, such as plasma-induced beam deflection, may also introduce small but non-negligible contributions.

Under non-ideal conditions, laser beams experience deflection during propagation in space plasma, resulting in a pointing eccentricity $\varepsilon$ and a deflection angle $\alpha$ at the receiver plane. As shown in Figure 2, the red cross marks the center of the laser beam when it reaches the receiver. The displacement between the center of the laser beam and the center of the receiver is defined as $\varepsilon$, and the angle difference between the final direction and the initial direction is defined as $\alpha$. Both parameters introduce tilt-to-length (TTL) coupling noise, and $\alpha$ is more directly relevant to TTL noise (Edwards et al., 2025; Zhao et al., 2020), which necessitates strict requirements for pointing stability in all such missions. Therefore, it is essential to quantitatively assess the values of $\varepsilon$ and $\alpha$ resulting from the plasma-induced light deflection for TQ's laser links.

## 3. Data and Method

### 3.1. MHD Simulation

The laser links of TQ are deployed in space around the Earth with an arm length of about $1.7 \times 10^5$ km, while the space is not a vacuum but is filled with plasma and the magnetic field. The pointing error caused by space plasma arises from the variation of the refractive index along the laser propagation link, and the refractive index is proportional to the plasma number density. Hence, to study the laser pointing error induced by space plasma, we need to obtain the plasma number density along the laser link first. Magnetohydrodynamics (MHD) simulation is widely used in space weather research (Groth et al., 2000; Jiang, 2024; Zhang et al., 2023). In this work, we use an MHD model, the Space Weather Modeling Framework (SWMF), to obtain the distribution of plasma around the laser propagation links on the TQ orbital plane (Tóth et al., 2005, 2012).

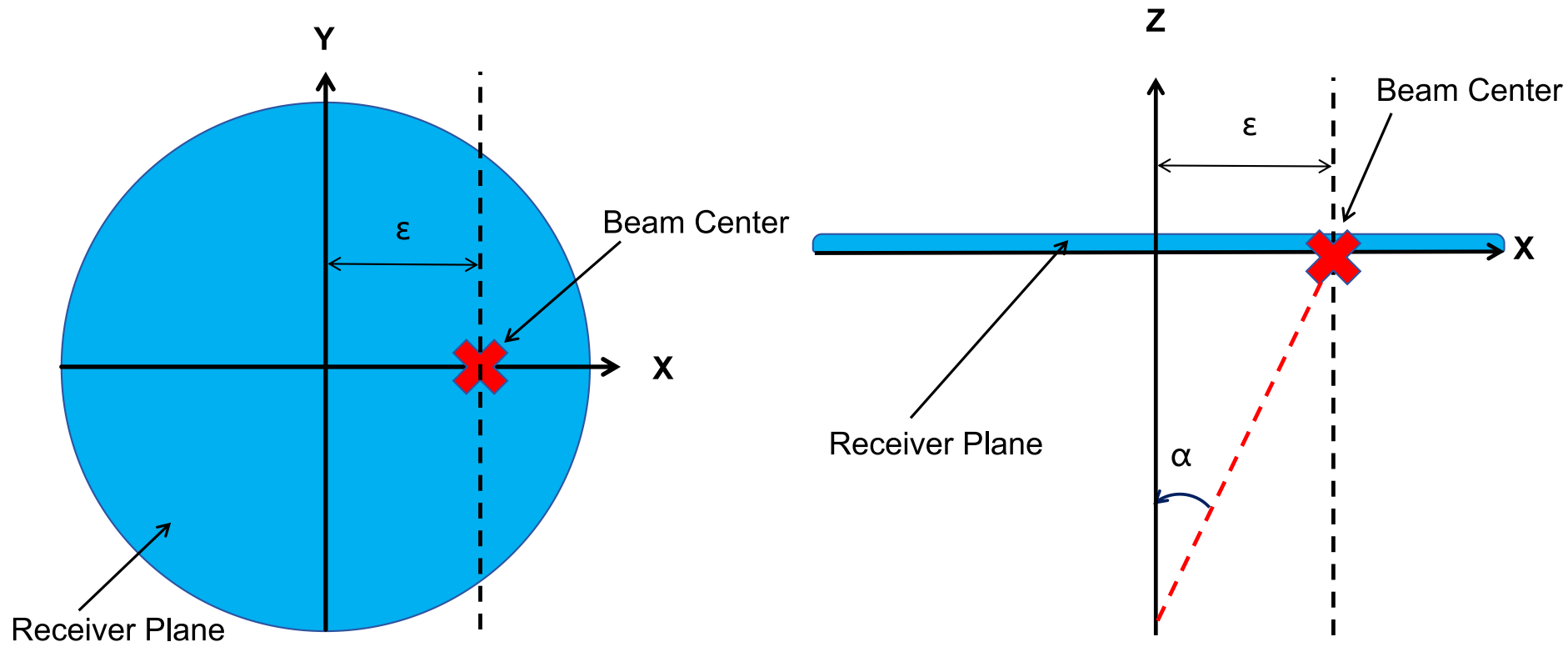

**Figure 2.** The schematic diagram of the top and front views of receiver plane, indicating the definition of $\alpha$ and $\varepsilon$. The red cross marks the center of the laser beam when it reaches the receiver after being deflected during propagation. $\varepsilon$ is defined as the transverse displacement between the beam center and the center of receiver plane, while $\alpha$ is defined as the final angular deviation of the beam direction compared with the original beam direction.

The SWMF is a publicly available model developed by the University of Michigan for simulating the space physics processes of the Sun, interplanetary medium, Earth, and other planetary space environments. This model can be accessed through NASA's Community Coordinated Modeling Center (CCMC) and has been widely proven and used (Jivani et al., 2023; Keebler et al., 2025; Samsonov et al., 2020; Welling et al., 2017). We use the global magnetosphere module of the SWMF to simulate the interaction between the solar wind and the Earth's magnetosphere. Based on the input solar wind data, we finally obtain plasma parameters near the TQ orbit, such as the bulk flow velocity, number density, and magnetic field. The temporal resolution of the simulation in this work is 1 min.

The solar wind dynamic pressure $(P_{\mathrm{dyn}})$ is a space environment parameter that has the highest linear correlation with laser propagation noise (Liu et al., 2024). Qualitatively, as $P_{\mathrm{dyn}}$ increases, the Earth's magnetosphere is further compressed (Parker, 1958; Samsonov et al., 2020; Shen et al., 2022; Zhou et al., 2018), leading to an increased plasma number density in the magnetosheath, and the refractive index and pointing error increase. The average $P_{\mathrm{dyn}}$ $(\overline{P}_{\mathrm{dyn}})$ from 1999 to 2021, which covers approximately 2 solar cycles, was $\overline{P}_{\mathrm{dyn}} \approx 2.04$ nPa. In this work, we selected a time interval that starts at 00:12 UT on 30 March 2017 and lasts one TQ orbital cycle (3.65 days), with a value of $\overline{P}_{\mathrm{dyn}} \approx 2.06$ nPa, which is close to the 23-year overall average value. The variations of solar wind parameters throughout the orbital cycle are shown in Figure 3. The zero point on the $x$-axis corresponds to the start time. From top to bottom, the first panel shows the variation of $P_{\mathrm{dyn}}$. The second panel shows the variation of plasma number density. The third and fourth panels show the variations of the magnetic field and solar wind velocity of three components, respectively. During this time interval, the value of $P_{\mathrm{dyn}}$ exhibits variations. It reaches a maximum value of over 3 nPa at approximately 0.4 TQ orbital cycles, and then drops to a minimum value of about 1 nPa at around 0.7 orbital cycles. Also, it is shown in Figure 3 that there is a strong correlation between the plasma number density and $P_{\mathrm{dyn}}$.

Taking the solar wind parameters in the selected time interval as input, we use SWMF to obtain the variation of space parameters around the orbit of TQ. Figure 4 shows the MHD simulation results at 00:12 UT on 30 March 2017. The color bar represents the space plasma number density. The TQ orbit is marked by the red circle, while S1, S2, and S3 denote the positions of the three satellites in the TQ constellation. Laser links are represented by colored lines. In the Geocentric Solar Ecliptic (GSE) coordinate, $\xi$ and $\zeta$ are denoted as follow: $\xi$ is the intersection line between the orbit plane and the ecliptic plane; $\zeta$ is perpendicular to the ecliptic and approximately parallel to the TQ orbital plane. Coordinates in Figure 4 are in units of Earth radius $R_{\mathrm{E}}$.

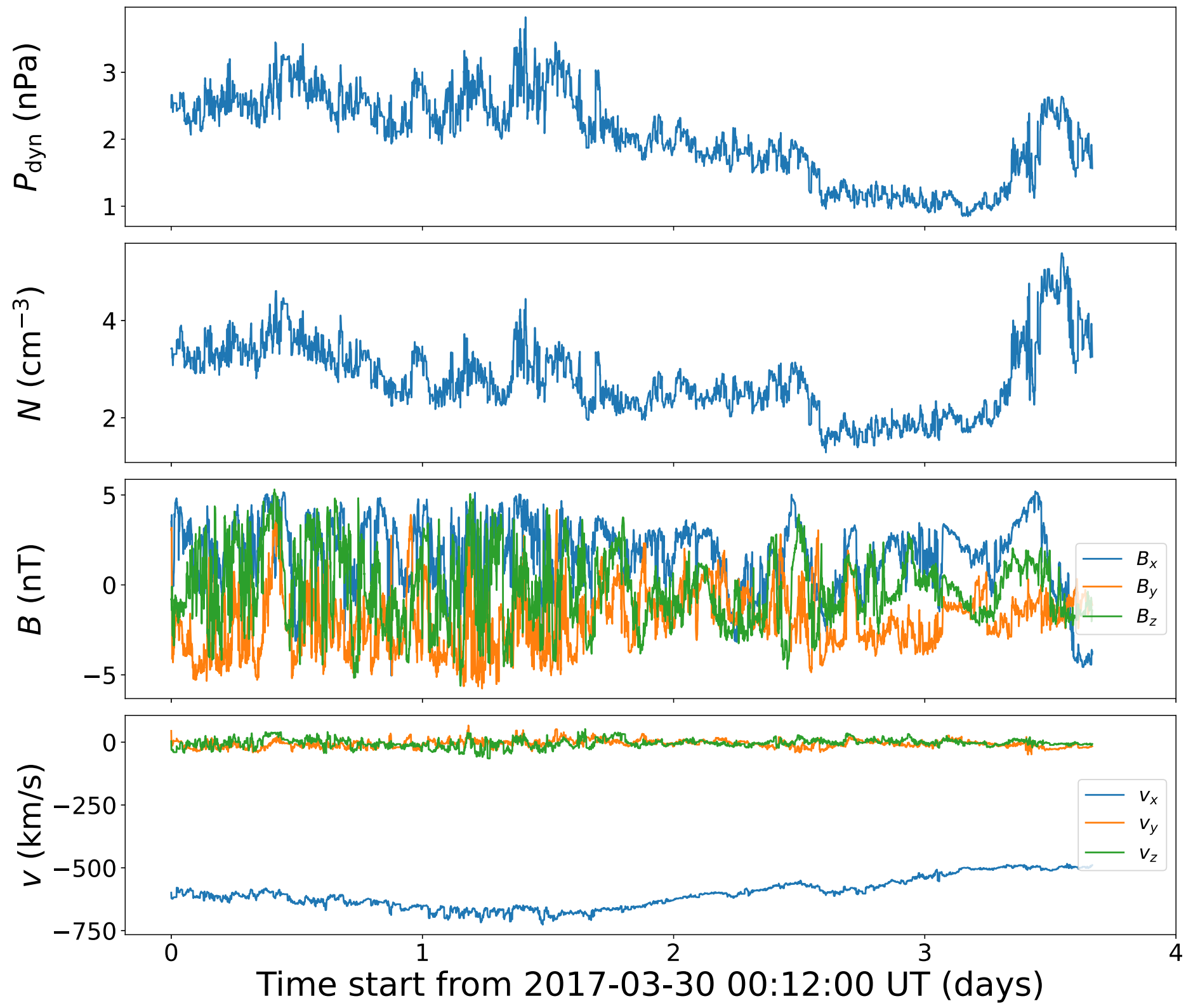

**Figure 3.** The temporal variations of solar wind parameters over one TQ orbital cycle starting from 00:12 UT on 30 March 2017. The lines represent the values of the corresponding solar wind parameters at each time point. From top to bottom: the first panel shows the $P_{dyn}$, the second panel shows the plasma number density, the third panel shows the magnetic field components, and the fourth panel shows the solar wind velocity components.

The left, middle, and right panels of Figure 4 show the MHD simulation results for $\varphi_s = 0°$, 45°, and 90°, respectively. For $\varphi_s = 0°$, the orbital plane of TQ is approximately vertical to the Sun-Earth line. At this point, the magnetosheath region (the area with a higher plasma number density) on the orbital plane exhibits a nearly ring shape, and the TQ orbit is also located within the magnetosheath at this moment. For $\varphi_s = 45°$ and 90°, Earth's magnetosphere structures are similar, with a distinct bow shock in the orbital plane. Under these conditions, the TQ orbit and the laser links pass through regions including the solar wind, bow shock, magnetosheath, magnetopause, and magnetotail.

To sum up, as $\varphi_s$ varies, the TQ orbit and the laser links are distributed across different regions of Earth's magnetosphere. As the satellites move along the TQ orbit, the laser links are jointly influenced by two factors: changes in plasma distribution due to the variations in solar wind interacting with the Earth's magnetosphere, and variations in the path position of the laser links caused by the orbital motion of the TQ constellation.

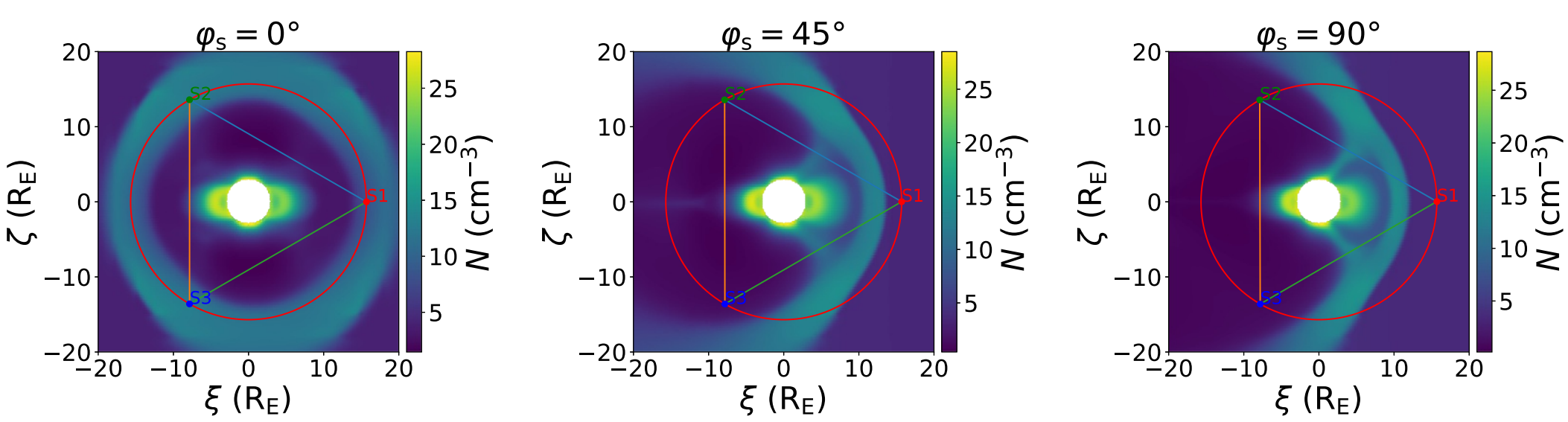

**Figure 4.** Distribution of plasma around TQ orbit for $\varphi_s = 0°$, 45° and 90° at 00:12 UT on 30 March 2017. $\xi$ is the intersection line between the orbit plane and the ecliptic, and $\zeta$ is perpendicular to the ecliptic. The red cycle is the orbit of TQ satellites. Coordinates are in GSE system and distances are in units of Earth radius $R_E$.

### 3.2. Ray-Tracing Method

Based on the distribution of refractive index around the laser links, the ray-tracing method can calculate the deflection of a laser beam propagating through space plasma. The refractive index distribution can be obtained according to the Appleton-Hartree (A-H) Equation. Since the space plasma can be considered collisionless, and the gyrofrequency is much lower than the plasma frequency, the A-H Equation can be simplified to,

$$\mu^2 = 1 - \frac{\omega_{\mathrm{p}}^2}{\omega^2} \tag{1}$$

where $\omega$ is the angular frequency of the diode-pumped Nd:YAG laser used for TQ, $\omega_{\mathrm{p}}$ is the plasma frequency. Thus, the refractive index μ is as follows,

$$\mu = \sqrt{1 - \frac{KN}{f^2}} \tag{2}$$

where $N$ is the plasma number density, and $f$ is the frequency of the laser used for TQ. The parameter $K$ can be deduced as $K = e^2/(4\pi^2 m\varepsilon_0) = 80.6\ \mathrm{m^3 s^{-2}}$, where $e$ is the elementary charge, $m$ is the electron mass, and $\varepsilon_0$ is vacuum permittivity. Thus, based on the plasma number density distribution obtained from the MHD simulation, the A-H Equation can be used to calculate the refractive index distribution around the laser links.

When light propagates through an inhomogeneous medium, it deflects if the direction of the light ray differs from the direction of the refractive index gradient of the medium at that location. The ray-tracing method can be used to study light deflection in inhomogeneous media (Feng et al., 2010; Hobiger et al., 2008).

The vector form of the differential equation of the light rays is shown as follows:

$$\frac{\mathrm{d}}{\mathrm{d}s}\left(n\frac{\mathrm{d}\mathbf{r}}{\mathrm{d}s}\right) = \nabla n \tag{3}$$

in which $\mathbf{r}$ is the position vector of the light, $n$ is the refractive index, and $s$ is the length of the arc of the ray.

Then Equation 3 can be deduced to the following first-order system of equations:

$$\begin{aligned} \frac{\mathrm{d}\mathbf{T}}{\mathrm{d}s} &= \nabla n \\ \frac{\mathrm{d}\mathbf{r}}{\mathrm{d}s} &= \frac{1}{n}\mathbf{T} \end{aligned} \tag{4}$$

where $\mathbf{r} = (x, y, z)$ is the position vector, $\mathbf{T} = (T_x, T_y, T_z) = n(\mathrm{d}x/\mathrm{d}s, \mathrm{d}y/\mathrm{d}s, \mathrm{d}z/\mathrm{d}s)$ is sometimes called the ray vector. Here, we use numerical methods to solve Equation 4 for the purpose of implementing ray-tracing. There are common ray-tracing methods, including the Euler method, the Taylor method, and the Runge-Kutta method. For a balance between accuracy and computational speed of ray-tracing, the fourth-order Runge-Kutta method is adopted in this work. The numerical solution via this method can be expressed as follows (Xiong et al., 2022).

$$\begin{aligned}\mathbf{r}_{n+1} &= \mathbf{r}_n + \frac{1}{6}h\left[\mathbf{T}_n + 2\left(\mathbf{T}_n + \frac{h}{2}\mathbf{D}(\mathbf{r}_n)\right)\right.\\ &\left. +2\left(\mathbf{T}_n + \frac{h}{2}\mathbf{D}\left(\mathbf{r}_n + \frac{h}{2}\mathbf{T}_n\right)\right) + \mathbf{T}_n + h\mathbf{D}\left(\mathbf{r}_n + \frac{h}{2}\mathbf{T}_n + \frac{h^2}{4}\mathbf{D}(\mathbf{r}_n)\right)\right]\\ \mathbf{T}_{n+1} &= \mathbf{T}_n + \frac{1}{6}h\left[\mathbf{D}(\mathbf{r}_n) + 2\mathbf{D}\left(\mathbf{r}_n + \frac{h}{2}\mathbf{T}_n\right)\right.\\ &\left. +2\mathbf{D}\left(\mathbf{r}_n + \frac{h}{2}\mathbf{T}_n + \frac{h^2}{4}\mathbf{D}(\mathbf{r}_n)\right) + \mathbf{D}\left(\mathbf{r}_n + h\mathbf{T}_n + \frac{h^2}{2}\mathbf{D}(\mathbf{r}_n)\right)\right]\end{aligned} \tag{5}$$

Here, the subscripts $n$ and $n + 1$ denote the ray-tracing step numbers. Under the two-dimensional conditions, $\mathbf{D}$ can be expressed as $\mathbf{D} = n(\partial n/\partial x, \partial n/\partial y)$. And $h$ is the optical step size and is defined as $h = \Delta s/n$, where $\Delta s$ is the spatial step size.

By inputting the initial position vector $\mathbf{r}_0$ and initial ray vector $\mathbf{T}_0$, the spatial step size $\Delta s$, as well as the spatial refractive index gradient distribution $\nabla n$, the fourth-order Runge–Kutta method given in Equation 5 can be employed to iteratively compute the position vector $\mathbf{r}$ and the ray vector $\mathbf{T}$ of the next step during the propagation of the ray.

Ultimately, the deflection angle $\alpha$ between $\mathbf{T}$ and $\mathbf{T_0}$ can be obtained. Meanwhile, the eccentricity of the light $\varepsilon$ can be determined from the displacement between the final position of the deflected ray and the originally targeted position.

## 4. Ray-Tracing Results

### 4.1. Ray-Tracing at a Single Moment

Taking 00:12 UT on 30 March 2017 as an example, the positions of TQ satellites S1, S2 and S3 at this moment are shown in Figure 4. S1, S2 and S3 are located at $(15.696\ R_\mathrm{E}, 0\ R_\mathrm{E})$, $(-7.848\ R_\mathrm{E}, 13.593\ R_\mathrm{E})$, and $(-7.848\ R_\mathrm{E}, -13.593\ R_\mathrm{E})$, respectively. The satellites approximately form an equilateral triangle. Here, we denote the laser link from S1 to S2 as L12. Similarly, we can label the other two links as L23 (from S2 to S3) and L31 (from S3 to S1). For a balance between the accuracy and computational speed of ray-tracing, we perform ray-tracing for these three laser links with a step size of $\Delta s = 0.2\ R_\mathrm{E}$.

According to the plasma number density obtained from MHD simulations, A-H Equation is used to calculate the refractive index on the orbital plane of TQ. We also calculate the gradient of the refractive index at each grid point in advance to facilitate the ray-tracing calculation. Since the position of each step of the ray may not fall exactly on a grid point during ray-tracing, an interpolation method is adopted. The relative difference in the results between the linear interpolation method and third-order spline interpolation is lower than 3%. For faster calculation speed, we use a linear interpolation method to interpolate both the refractive index $n$ and its gradient $\nabla n$ at each step location along the ray. Thus, we obtain the change in the deflection angle d$\alpha$ from step $n$ to $n + 1$.

Figure 5 displays the results of the ray-tracing simulation at 00:12 UT on 30 March 2017. The $x$-axis represents the normalized arm length of the laser link. The top panels of Figure 5 show the plasma number density along the laser link. It shows that the plasma number density distribution along the laser links of the TQ satellites exhibits noticeable variations for different values of $\varphi_\mathrm{s}$. For $\varphi_\mathrm{s} = 0°$, the plasma density is higher near both ends of the links and lower in the middle region. Comparing this condition with Figure 4, it is indicated that this occurs because all three satellites are located within the high-density magnetosheath. At the same time, the laser paths pass through some relatively low-density areas of the magnetosphere. For $\varphi_\mathrm{s} = 45°$ and 90°, the plasma density along L12 and L31 increases significantly when the links cross the magnetosheath and decreases to lower levels when passing through the inner magnetosphere and solar wind regions. In contrast, link L23 resides entirely within the inner magnetosphere, resulting in a consistently low plasma density along its path.

The middle panels of Figure 5 represent the cumulative deflection angle $\alpha$ at each step relative to the initial laser direction. These panels also represent $\varepsilon$, which shows the distance between the local beam center and the beam center that maintains its original direction. The values of $\alpha$ for all three angles are on the order of $10^{-21}$ rad. The values of $\varepsilon$ are on the order of picometers and change according to the local $\alpha$ values. The bottom panels of

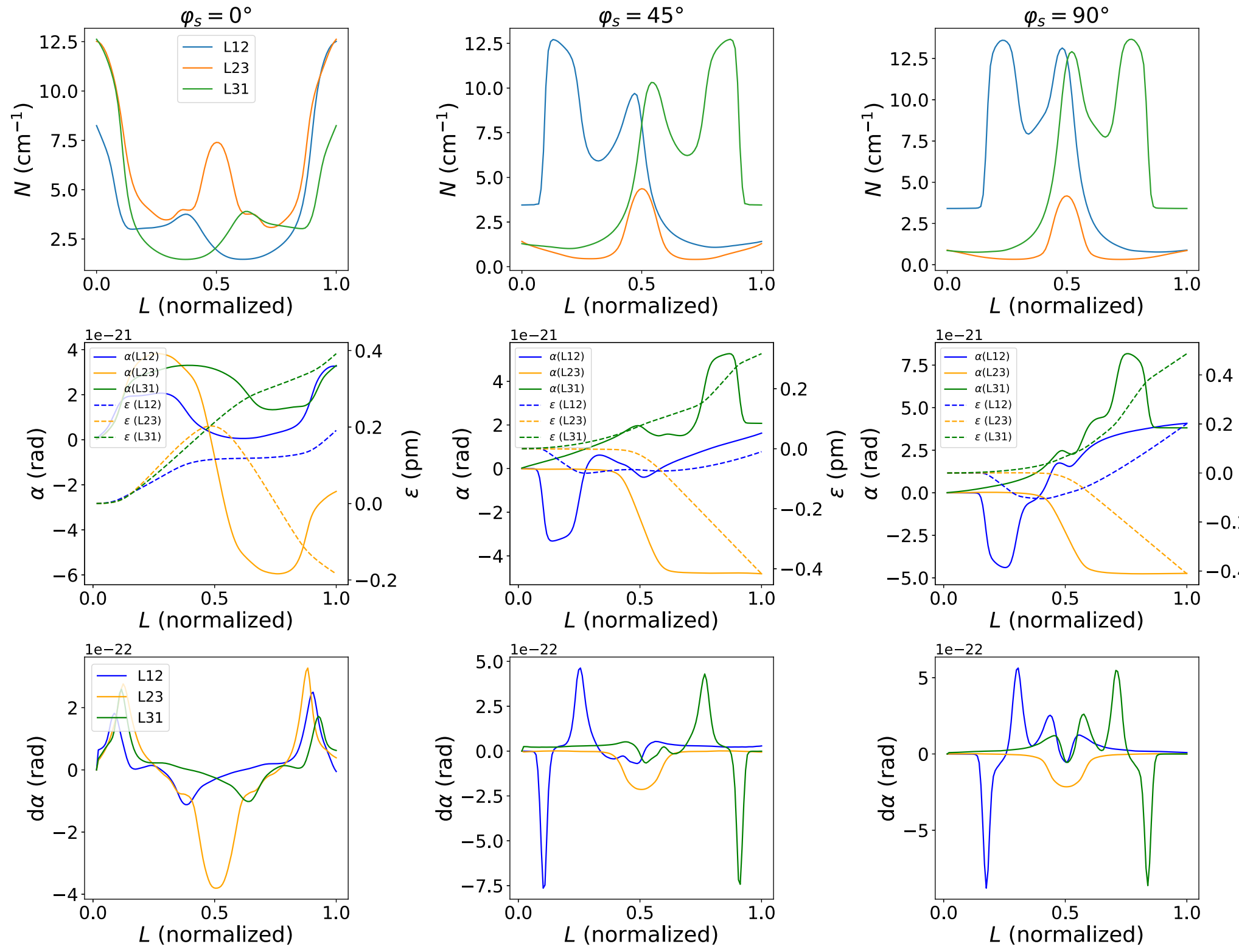

**Figure 5.** The top panels are the plasma number density distribution along the TQ inter-satellite laser link. The middle panels are the variation of the laser deflection angle $\alpha$ during the propagation along the laser link using the fourth-order Runge-Kutta method, and the pointing eccentricity $\varepsilon$ is represented by the dashed line for each laser link, respectively. The bottom panels are the incremental change d$\alpha$ of the deflection angle per step computed along the laser link.

Figure 5 represent the angle change d$\alpha$ at each step, which shows that the values of d$\alpha$ are generally around $10^{-22}$ rad. For $\varphi_s = 45°$ and 90°, the maximum values can exceed $5 \times 10^{-22}$ rad.

Moreover, we note that due to the roughly symmetric plasma density distribution along links L12 and L23, the variation of d$\alpha$ with normalized arm length also exhibits noticeable symmetry in the bottom panels. This condition is consistent with the spatial correspondence between the laser links and plasma structures shown in Figure 4.

### 4.2. Ray-Tracing During a TQ Orbital Cycle

In Section 4.1, the analysis above only reflects the condition at a specific time with a specific satellite location. In reality, as the TQ satellite rotates along its orbit, its spatial orientation relative to the Earth's magnetosphere also changes continuously. This occurs because the satellite rotates within its orbital plane, while the magnetosphere remains relatively fixed in orientation relative to the Sun–Earth line during one orbital cycle of TQ. Furthermore, solar wind varies over time, leading to corresponding temporal changes in plasma distribution. Therefore, it is essential to conduct a time-dependent analysis of the plasma-induced errors over an entire orbital cycle to assess these dynamic effects.

Based on the MHD simulation with a temporal resolution of 60 s, we perform ray-tracing calculations at one-minute intervals. After taking the orbital motion of the TQ satellites into account, we obtain the temporal variations of both the laser deflection angle $\alpha$ and the pointing eccentricity $\varepsilon$, as shown in Figure 6.

In Figure 6, the horizontal axis represents the normalized TQ orbital cycle, and the vertical axis in the top panels and bottom panels denotes the pointing eccentricity $\varepsilon$ and the deflection angle $\alpha$ of the laser beam upon arrival at the receiving satellite, respectively.

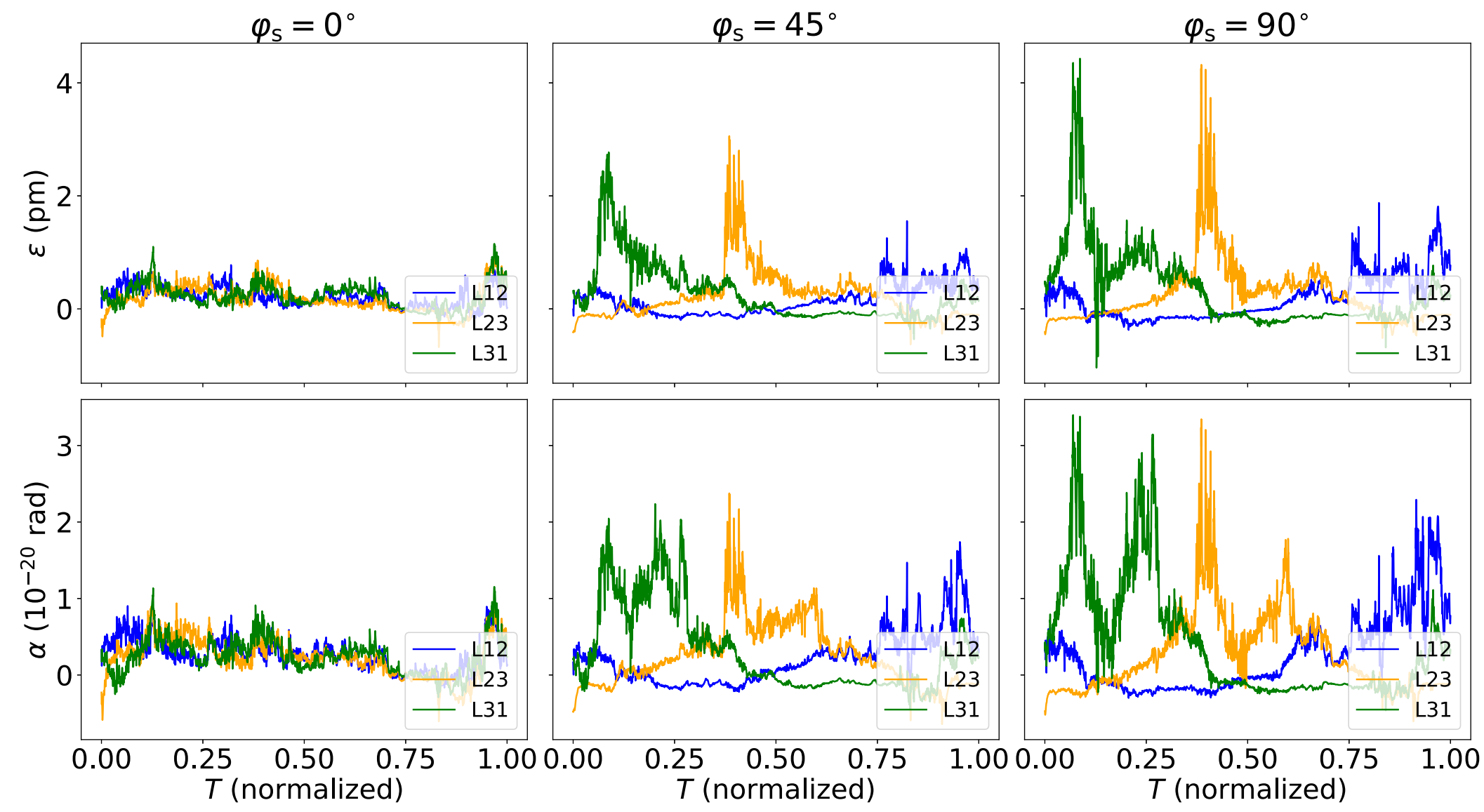

**Figure 6.** Variation of the pointing eccentricity $\varepsilon$ (top panels) and final deflection angle $\alpha$ (bottom panels) over one TQ orbital cycle, starting from 00:12 UT on 30 March 2017, for the three laser links at $\varphi_s = 0°$, 45° and 90°. The horizontal axis represents the normalized TQ orbital cycle.

In the top panels of Figure 6, we note that for $\varphi_s = 0°$, the values of $\varepsilon$ for all three links are generally similar at any given time. In contrast, for $\varphi_s = 45°$ and 90°, significant differences occur in the temporal evolution of $\varepsilon$ for the three laser links. Taking $\varphi_s = 45°$ as an example, the absolute value of $\varepsilon$ for L31 reaches its maximum value of approximately 3 pm at around 0.1 TQ orbital cycle, while the absolute value of $\varepsilon$ for L12 and L23 remain below 1 pm at that time. Similarly, at around 0.4 orbital cycle, the absolute peak of $\varepsilon$ for L23 is at about 3 pm, while L12 and L31 both remain below 1 pm. The phenomenon occurs because, as the satellite moves, when the laser link passes through the magnetosheath at angles relative to the plasma number density gradient becomes large, $\varepsilon$ increases. At this moment, the other two laser links either pass through the magnetosheath at a smaller angle to the plasma number density gradient or do not pass through magnetosheath. It indicates that the orientation of the laser links can influence $\varepsilon$. When L12 is at approximately 0.7–0.8 TQ orbital cycle, the absolute value of $\varepsilon$ is about 1 pm, which is lower than that of L31. This is because during this cycle, the $P_{dyn}$ is only about 1 Pa, leading to a reduction in the density of the magnetosheath, the corresponding solar wind parameters are shown in Figure 3.

As a result, the maximum $\varepsilon$ of L12 does not reach the same level as those of L23 and L31. This difference indicates that the time-varying of $P_{dyn}$ can affect the pointing error of TQ.

In the bottom panels of Figure 6, similar to the top panels, the values of $\alpha$ for all three links remain relatively consistent over most of the time for $\varphi_s = 0°$. Whereas, for $\varphi_s = 45°$ and 90°, the temporal variation of $\alpha$ shows certain similarities with that of $\varepsilon$ described above. That is, the absolute value of $\alpha$ for L31 and L23 is around 0.1 and 0.4 orbital cycles, respectively. The absolute value of $\alpha$ reaches approximately $2 \times 10^{-20}$ rad and $3 \times 10^{-20}$ rad for $\varphi_s = 45°$ and 90°, respectively.

However, two additional peaks in $\alpha$ form around 0.2 and 0.6 orbital cycle, which are absent in the panel of $\varepsilon$. For $\varphi_s = 45°$, the amplitude of the peaks in L31 and L23 is about $2 \times 10^{-20}$ rad and $1 \times 10^{-20}$ rad, respectively. For $\varphi_s = 90°$, the amplitude is approximately $3 \times 10^{-20}$ rad and $2 \times 10^{-20}$ rad, respectively. This clear contrast between the prominent peaks in $\alpha$ and the lack of corresponding features in $\varepsilon$ indicates that under certain orbital conditions, the laser deflection angle becomes significant while the pointing eccentricity remains relatively small. This phenomenon may be attributed to the spatial relationship between the inter-satellite laser links and the magnetosphere. For example, there are pronounced deflections occurring near the endpoints of the link, which obviously influence $\alpha$ but have a relatively minor impact on $\varepsilon$. Such effects can lead to a divergence between the trends of $\alpha$ and $\varepsilon$ at those times.

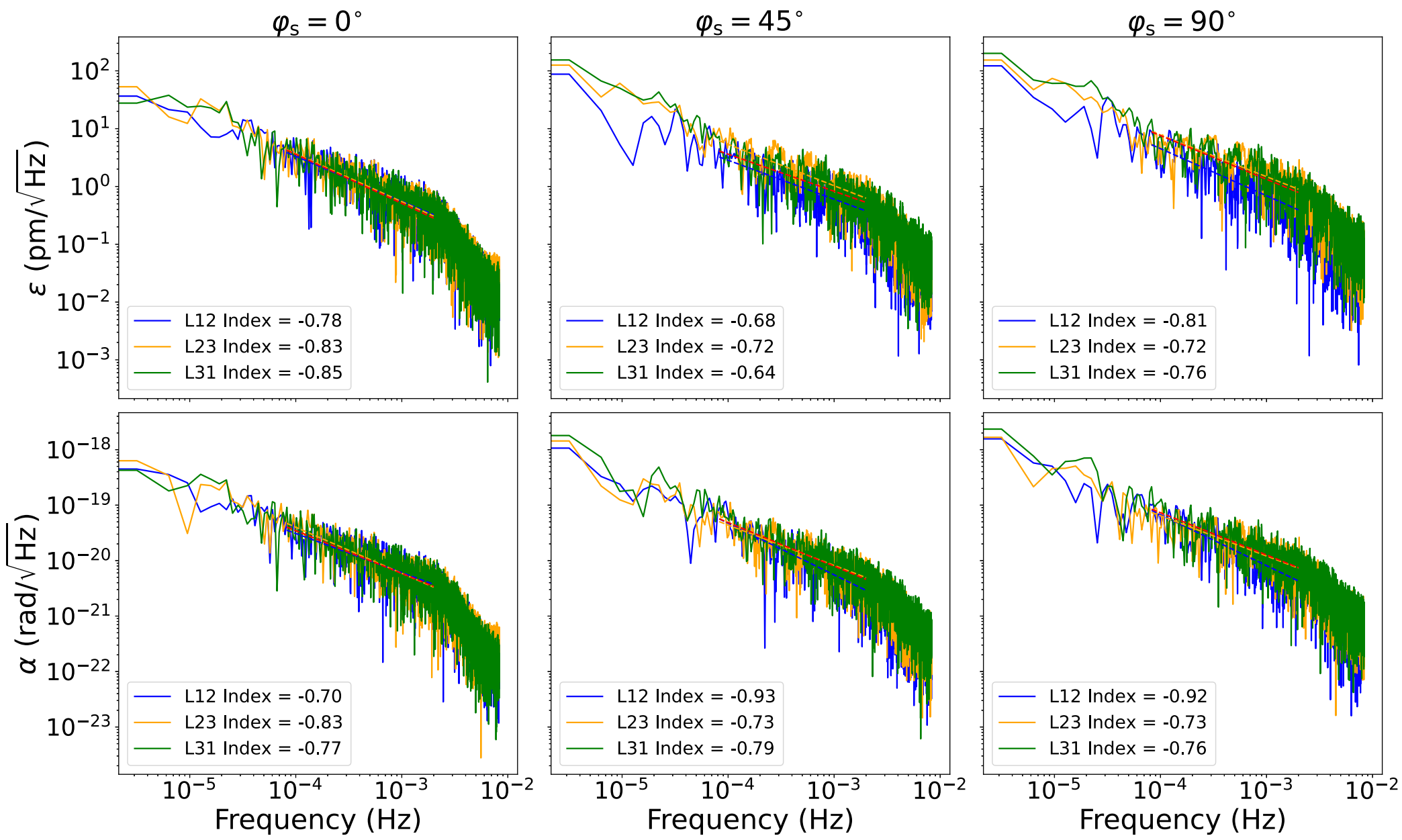

**Figure 7.** ASDs of the pointing eccentricity $\varepsilon$ (top panels) and deflection angle $\alpha$ (bottom panels) over one TQ orbital cycle, starting from 00:12 UT on 30 March 2017, for the three laser links at $\varphi_s = 0°, 45°, 90°$. The horizontal axis represents frequency. The dashed line represents the spectral index of each laser link for different $\varphi_s$, respectively.

In summary, during the orbital cycle, the influence of plasma on the laser deflection angle $\alpha$ is on the order of $10^{-20}$ rad, while the effect on the pointing eccentricity $\varepsilon$ is on the order of pm. The values of $\alpha$ and $\varepsilon$ also vary noticeably among the three laser links depending on the values of $\varphi_s$, $P_{dyn}$, and the orbital positions of TQ satellites.

## 5. Discussion

According to the time distribution of $\alpha$ and $\varepsilon$ in the TQ cycle, we calculate the amplitude spectral density (ASD) of $\alpha$ and $\varepsilon$ for $\varphi_s = 0°$, $45°$, and $90°$. And the ASDs results of $\alpha$ and $\varepsilon$ are shown in Figure 7.

As shown in Figure 7, the ASDs of $\alpha$ are about $1 \times 10^{-20}$ rad, $1.5 \times 10^{-20}$ rad, $2 \times 10^{-20}$ rad at 1 mHz for $\varphi_s = 0°, 45°$, and $90°$, respectively; $\varepsilon$ is approximately 0.9 pm, 1.6 pm, and 2 pm at 1 mHz for $\varphi_s = 0°, 45°$, and $90°$, respectively. We note that the ASD curves for both parameters exhibit similar characteristics. At $\varphi_s = 0°$, the ASD curves of the three links are highly consistent, reflecting the symmetric plasma distribution at this angle.

Meanwhile, we notice that as the frequency increases, $\alpha$ and $\varepsilon$ drop continuously. In the TQ orbit cycle, the spectral index for each laser link varies from $-0.93$ to $-0.64$, which is close to half of the power spectrum density (PSD) of the space plasma, with a value of about $-5/3$ (Huang et al., 2014; Tu & Marsch, 1995). This indicates that the deflection angle noise caused by plasma decreases with increasing frequency, a result that is consistent with the behavior of the optical path difference (OPD) noise. These findings suggest that the light deflection noise and OPD noise share a common physical origin, that is, turbulence in the space plasma (Bruno & Carbone, 2013; Huang et al., 2021; Wu et al., 2025).

Moreover, our ray-tracing method can be adopted to compute the optical beam deflection effect resulting from laser propagation for other space-based GW missions, such as LISA and Taiji. Both of LISA and Taiji are heliocentric GW detectors, the orbits of LISA and TJ are both in the solar wind, where there are no structures such as the magnetosheath or magnetosphere. Therefore, unlike the geocentric GW detector TQ, the laser links of LISA and TJ will not pass through the areas with high density gradients such as bow shock waves and magnetopause. Thus, for heliocentric GW detectors, the deflection of the laser in the space plasma may be smaller than that for the geocentric GW detector.

Here, we study the deflection of laser propagation in space plasma. In a case where $P_{dyn}$ is close to the average value during 2 solar cycles, the deflection angle $\alpha$ caused by space plasma is on the order of $10^{-20}$ rad for TQ, and the resulting pointing eccentricity $\varepsilon$ is approximately on the order of pm. Compared to TQ requirements for pointing stability ($\sim$ 10 nrad) and pointing jitter ($\leq$10 nrad/$\sqrt{\text{Hz}}$), the light deflection induced by space plasma is relatively weak. This suggests that the impact of space plasma on TQ's TTL coupling noise is expected to be manageable.

While this study demonstrates the negligible impact of space plasma under quiet to moderate conditions, it should be noted that the present study has several limitations that should be considered when interpreting the results. First, we use the Space Weather Modeling Framework to simulate the spatial distribution and temporal evolution during one TQ orbital cycle (3.65 day). Due to the limitations of the resolution in the simulation model, this model cannot resolve small-scale physical processes, such as small-scale turbulence, and so on. Thus, the ray-tracing based on this model in this work cannot obtain the light curve due to small-scale processes, for example, turbulence. In addition, the analysis is conducted under quiet to moderate space weather conditions. Although the strength of space weather is usually at this stage, extreme space weather events are not fully explored. For extreme cases, for example strong coronal mass ejections, geomagnetic storms, and substorms, the magnetosphere can be strongly compressed. As a result, the gradient of the plasma density near the magnetopause boundary layer and the bow shock becomes much larger than that for quiet or moderate conditions. Consequently, the deflection angle $\alpha$ and the pointing eccentricity $\varepsilon$ during laser propagation may increase significantly. The SWMF is widely used (Haiducek et al., 2017; Liemohn et al., 2018) and has been validated by in situ observations (Kwagala et al., 2020; Wang et al., 2008). In this work, the interaction between the solar wind and magnetosphere, and the structure of the magnetosphere is well revealed by the SWMF. While, the fine physical processes, such as Kelvin-Helmholtz instability (KHI) (Faganello & Califano, 2017; Kavosi & Raeder, 2015; Li et al., 2012; Li et al., 2016), eruptions of plasma from the inner magnetosphere (Bagby-Wright et al., 2023; Darrouzet et al., 2006), are not simulated in this work. Thus, the deflection angle due to the fine structures, such as KHI of the magnetopause boundary layer and plasmaspheric plumes, cannot be revealed in this work. Due to the aforementioned constraints, plasma-induced pointing errors remain negligible for TQ under quiet to moderate space weather conditions.

## 6. Conclusion

Space plasma has a certain impact on the TQ laser links when the laser links pass through the space plasma (Jennrich et al., 2021; Su et al., 2021). In this work, we use an MHD model, SWMF, to obtain the distribution of plasma around the laser links on the TQ orbital plane under $\varphi_s = 0°, 45°$, and $90°$, respectively. The time interval of the simulation is one TQ orbital cycle, which lasts 3.65 days. The average $P_{dyn}$ during the simulation time interval is about 2.06 Pa, which closely matches the 23-year (over 2 solar cycles) average, 2.04 Pa. We consider this interval to be a moderate condition and a representative space weather environment. Based on the plasma number density obtained from MHD simulations, the Appleton–Hartree equation is used to derive the refractive index around the laser links. Subsequently, we applied the fourth-order Runge–Kutta method to compute the laser deflection angle $\alpha$ and pointing eccentricity $\varepsilon$ induced by space plasma over one full orbital cycle. During the orbital cycle, the variation of $\alpha$ is on the order of $10^{-20}$ rad under moderate conditions. The variation of $\varepsilon$ is on the order of pm. Under moderate conditions, the maximum values of $\alpha$ for the three links increase as the $\varphi_s$ increases. Similarly, the maximum values of $\varepsilon$ for the three links also increase as the $\varphi_s$ increases under such circumstances.

In conclusion, compared with the TQ pointing stability requirements ($\sim$ 10 nrad) and TQ pointing jitter requirements ($\leq$10 nrad/$\sqrt{\text{Hz}}$), the laser deflection induced by space plasma is several orders of magnitude weaker in the context of quiet to moderate space weather. Therefore, we conclude that the laser deflection caused by space plasma does not represent a fundamental risk to the TQ mission.

## Conflict of Interest

The authors declare no conflicts of interest relevant to this study.

**Acknowledgments**
The MHD simulations used in this work are provided by the Community Coordinated Modeling Center (CCMC) at Goddard Space Flight Center through the request system. This work is carried out using the SWMF tool developed at the University of Michigan's Center for Space Environment Modeling. S. W. is supported by NSFC (Grant 12473060 and 122261131504), and the National Key R&D Program of China (No. 2020YFC2201200).

## Availability Statement

All data used in this study are freely available online. The simulations used in this work are provided by the Community Coordinated Modeling Center (CCMC) at Goddard Space Flight Center through the request system (https://ccmc.gsfc.nasa.gov/, run number: JiXiang_Zhang_042923_1). The solar wind data are from NASA's OMNIWeb service (https://omniweb.gsfc.nasa.gov/). The core code associated with this manuscript has been deposited at Zenodo via https://doi.org/10.5281/zenodo.18632837 (S. Zhou & Su, 2026).